# Cu, Pu and Fe high $T_c$ superconductors: all the same mechanism!

Peter Wachter
Laboratorium für Festkörperphysik, ETH Zürich, 8093 Zürich, Switzerland

The new iron based high $T_c$ superconductors with $T_c$ up to 55 K have stirred new interest in this field. It is consensus that the BCS mechanism is not able to explain the high $T_c$'s. In the following we propose that spin holes in anti - ferromagnetic clusters combine to make nonmagnetic bipolarons, which can condense and lead to superconductivity.

The discovery of high $T_c$ superconductivity in iron based $Ln(O_{1-x}F_x)FeAs$ (Ln = La, Ce, Pr, Nd, Sm, Gd) with $T_c$ from 25 – 55 K has opened a new era of research [1-6] with materials, which are especially simple to prepare. Single crystals with $T_c$ of 53 K have been obtained for the Sm compound when grown under hydrostatic pressure of 30 kbar and T = 1350 - 1450 C with x = 0.2 [7].

The materials are layer structured and resemble thus the Cu high $T_c$ super-conductors, although two-dimensionality is not a prerequisite for Cu high $T_c$ superconductivity, inasmuch as a three-dimensional cubic cuprate exists in ceramic form with a $T_c$ of 117 K [8].

Existing comments on the discovery of the Fe superconductors [9.10] express their surprise that iron, being magnetic, does not destroy superconductivity. Fe, Ni and Co are the most well known ferromagnetic metals and there are general statements that ferromagnetism and superconductivity do not mix. Apparently these authors are not aware that even iron under high pressure becomes a superconductor. [11]. Also the boro - carbides $LnNi_2B_2C$ (Ln = lanthanides) contain Ni and are superconductors, in the case of Lu with 16 K [12].

It is also stated that with the iron compounds one has a second group of high $T_c$ superconductors after the cuprates. This leaves out the group of high $T_c$ plutonium superconductors. $PuCoGa_5$ has with 18.5 K a $T_c$, which may not lead to excitement, but in the field of actinides, where superconductivity is restrained to at most 3 K, the enhancement of $T_c$ is of equal size as from 23 K in pre-cuprate times to 136 K for the highest $T_c$ of a copper compound. In addition the presence of nonmagnetic Co in the Pu compound does not hamper superconductivity [13].

The key in understanding superconductivity with 3d, 4f and 5f elements lies in the fact that the ions can have magnetic or nonmagnetic configurations and they can be itinerant or localized in the solid.

There are thus **three** systems, which now exhibit high $T_c$ superconductivity and it has been shown, that the Pu compounds have the same superconducting mechanism as the cuprates [13]. We will show below that also the new iron superconductors have the same mechanism for superconductivity as the two other systems.

The essential feature in high $T_c$ superconductors is the fact that the parent compounds to the superconductors, i.e. the undoped compounds are antiferromagnets. Thus $La_2CuO_4$ with $T_N$ = 250 K [14] and $YBa_2Cu_3O_{6.5}$ with $T_N$ = 400K [15] contain only divalent $Cu^{2+}$ in the $3d^9$ configuration with spin ½. $PuCoGa_5$ is an intermetallic alloy with no long-range antiferromagnetism but it is superconducting with $T_c$ = 18.5 K [16]. $Pu^{3+}$ is in a magnetic J = 5/2 state with a magnetic moment corresponding to this configuration. Co and Rh in $PuRhGa_5$ ($T_c$ = 8.7 K) are in a nonmagnetic trivalent $3d^6$ configuration with the $t_2$ band filled completely. $PuCoGa_5$ exhibits above $T_c$ a paramagnetic Curie-Weiss law with a negative Curie temperature of $\theta_p$ = –3 K, indicative of antiferromagnetic fluctuations or short - range order, such as clusters etc. Iron in undoped LnOFeAs is in a magnetic state and exhibits antiferromagnetism or a commensurate SDW below 140 K. Mössbauer effect [17] and neutron scattering [18] yield a magnetic moment between 0.25 - 0.35 $\mu_B$.

In contrast to ferromagnetism, antiferromagnetism is no antagonist to superconductivity [19]. And in fact $Ce(Rh,Ir)In_5$ and others are simultaneously antiferromagnets and low $T_c$ superconductors. So if the high $T_c$ superconductors above want to be antiferromagnets they do not have to loose this property upon doping. But they do, and in fact there is no long-range antiferromagnetism in the doped superconducting compounds. So it is the doping itself, which destroys antiferromagnetism.

Now all antiferromagnets have only short - range correlations (but long range order). Thus if one substitutes part of the magnetic ions with nonmagnetic ions between about 5 and 20 %, long range antiferromagnetism will break down. This is the case in the doped superconducting compositions. Instead of antiferromagnetic gaps only pseudo - gaps remain [13] due to short - range order.

In the case of the cuprates the magnetic ions are the divalent Cu ions in the $3d^9$ configuration and doping e.g. $La_2CuO_4$ with Sr results in some trivalent $Cu^{3+}$ $3d^8$ ions, and in the given crystal field this configuration is nonmagnetic. This has already been suggested in the very first paper by Bednorz and Müller [20] and the copper in the superconductors is thus in a **mixed valence configuration**. Indeed $NaCu^{3+}O_2$ is a diamagnetic semiconductor [13]. The same is true in the classical 123 Cu compound $YBa_2Cu_3O_7$, which can be written as $Y^{3+}Ba_2^{2+}Cu_1^{3+}Cu_2^{2+}O_7^{2-}$ (overdoped, twinned SC, $T_c \approx$ 90 K, 33 % $Cu^{3+}$). In the n - doped cuprate $Nd_{2-x}Ce_xCuO_4$ the nonmagnetic copper is $3d^{10}$ $Cu^{1+}$, and the compound should be written as $Nd_{2-x}^{3+}Ce_2^{4+}Cu_{1-x}^{2+}Cu_x^{1+}O_4^{2-}$. In the Pu alloy also Pu can have two valences: the magnetic $Pu^{3+}$ $5f^5$ J = 5/2 and the nonmagnetic $Pu^{2+}$ $5f^6$ with J = 0. $Pu^{2+}$ is in the same configuration as $Sm^{2+}$, its 4f counterpart. High - resolution photoemission indeed yields two 5f configurations [21]. So again the Pu alloy is in a **mixed valence configuration [13].**

Iron in the new high $T_c$ superconductors could also be present in two valences, $3d^5$ as high spin and $3d^6$ as low spin. But it is not so! When one has an iron compound the Mössbauer effect is the first choice. The Mössbauer effect has two properties, namely the hyperfine splitting of the Mössbauer line when the material shows long - range magnetic order and the isomer shift, which measures the density of s electrons in the nucleus. Different d electrons in different valences screen the core charge differently and when the different valences, e.g. $3d^5$ and $3d^6$ are in the same compound then two isomer lines should be observed. This is not the case [17]. Instead only one singlet isomer line is observed, above 150 K in the same position for the doped and undoped compound. In the doped and superconducting material the position of the singlet isomer line exhibits only the normal and weak temperature variation, whereas the undoped and antiferromagnetic parent material exhibits the standard hyperfine splitting into 6 lines as demanded by the core multiplet when cooling near or below $T_N$. In other words the superconductor has only one singlet isomer line, above and below $T_c$. There is thus **only one iron valence.**

But definitely, the doping must introduce a certain amount of nonmagnetic iron to destroy long - range antiferromagnetism, without changing the valence. This is possible and is the new idea. Replacing oxygen with fluorine not only introduces electrons, but it changes locally the crystal field acting on the iron ions. The spin configuration of an ion in a solid depends on the crystal field. If we assume the iron is divalent $3d^6$ a high spin configuration is $t_2^4 e^2$ in a $\Gamma_5$ configuration and a low spin configuration is $t_2^6$ in a nonmagnetic $\Gamma_1$ state. So with the **same valence we can have a magnetic and nonmagnetic configuration**, triggered by variation of the local crystal field as induced by doping. This can be illustrated in Fig. 1. So we can have for all 3 high $T_c$ superconducting systems the same mechanism where through doping we introduce nonmagnetic states in the otherwise antiferromagnetic matrix and thus destroy long – range magnetic interaction. But we retain short - range antiferromagnetism in clusters or fluctuations. This is in agreement with all experimental observations.

But is iron really divalent in the new high $T_c$ superconductors? In Ref. 17 there is an explicit statement that iron is divalent as deduced from the isomer shift of the Mössbauer effect which is typically S = 0.52 mm/sec. Generally speaking divalent iron compounds such as $FeCl_2$ etc. have their isomer shifts between 0.85 – 1.0 mm/sec, trivalent iron compounds between 0.0 and 0.1 mm/sec. The measured isomer shift of S = 0.52 mm/sec is not very helpfully exact in between, but exactly at the position of divalent FeS. Covalency has also an influence on the isomer shift but it is difficult to estimate this effect on the iron superconductors. However, iron could also be intermediate valent, i.e. a quantum mechanically hybridization between di - and trivalent iron, as is well known in rare earth and actinide compounds [22]. Then there would only be one isomer line as in gold $Sm^{2.8+}S$ and strong moment quenching and antiferromagnetism is an option as in $Tm^{2.7+}Se$. Also $Fe^xSi_2$ is such an intermediate valent compound. But at the moment it is too early to make a certain decision.

Band structure calculations [23, 24] on the new iron compounds indeed find also divalent iron with $3d^6$ configuration, which fill a $dt_2$ band completely, sitting exactly at and below the Fermi edge $E_F$. Above $E_F$ there is a gap in the density of states with about 0.5 eV width, before empty de bands start. The filled $dt_2$ band extends from about $E_F$ until –2 eV, whereas O-p and As-p bands are between –2 and –6 eV. The problem is that these band structure calculations find no support from experimental density of states photoemission studies [25], where only **250 meV** below $E_F$ a minute peak is observable, then a plateau which rises slowly towards larger binding energy, until a peak is reached at –10 eV. Theoretically, also the undoped material is nonmagnetic because a filled $dt_2$ band cannot be magnetic and a possible spin density wave is commensurate [17]. Experimentally, however, the materials are antiferromagnets.

The problem lies in the question are the d states of iron localized or itinerant? Starting the band structure calculations with plane wave pseudo – potentials [23, 24] will necessarily result in bands for all electronic states. This is the same shortcoming as has been made by Mattheiss [26] in the first band structure calculation on $La_2CuO_4$ who found itinerant Cu 3d bands, whereas experimentally it was realized that Cu $3d^9$ is localized and the material is an insulator [27]. At the moment a bad metallic conductivity for the iron compounds can be explained with extrinsic carriers due to imperfections of the polycrystalline materials. We will assume that the iron $3d^6$ states are quasi localized near $E_F$, yielding the minute density of states peak near $E_F$ in the photoemission spectrum [25] These localized $3d^6$ states lead to antiferromagnetism in the undoped material.

Now a spin hole such as $Cu^{3+}$ or $Cu^{1+}$ or low spin $Fe^{2+}$ or nonmagnetic $Pu^{2+}$ (J=0) in an antiferromagnetic two – dimensional cluster acts as a small ferromagnetic region, termed a magnetic polaron by Nagaev [28] or later even a ferron [29]. This is shown in Fig. 2a for 2 independent spin holes in a two – dimensional antiferromagnetic cluster. It has been shown experimentally that one can observe these ferromagnetic spin holes by their individual rotation of the optical polarization plane of linearly polarized light due to the Kerr effect [13]. A simple antiferromagnetic cluster does not have a net magnetization and cannot cause a rotation of the polarization.

There is a mutual magnetic attraction between the magnetic polarons or ferrons to form ferromagnetic bipolarons (Fig. 2b) or nonmagnetic bipolarons (Fig. 2c) which then can make a Bose condensation at $T_c$ and cause superconductivity. This has been predicted by Alexandrov et al. [30] and Mott [31]. One single magnetic polaron as in Fig. 2a breaks 4 magnetic bonds to the nearest neighbours; two independent magnetic polarons break 8 magnetic bonds. Only 7 magnetic bonds are broken for a ferromagnetic (triplet) magnetic bipolaron as in Fig. 2b, but only 6 magnetic bonds are broken for a nonmagnetic bipolaron (Fig. 2c) The net effect is a binding energy due to a gain in magnetic exchange energy proportional to $T_c$.

Now we should discuss how the magnetic polaron comes to its charge, since in Fig. 2a all ions shown, with or without spin, are $3d^6$ ions. When the chemical formula $LnO_{1-x}F_xFeAs$ is correct we have iron and Ln planes surrounded each tetrahedrally with 4 As or 4 O ions, respectively. These planes are charged, $(FeAs)^-$ and $(LnO)^+$ [24] and make sort of an ionic bonding between the planes. Replacing oxygen with F, i.e. an ion with another valence, must induce also another valence in another ion in the lower energy As band. Since there is a sizable hybridization between iron d states and the As p band, the iron next to this As ion with different valence will feel locally a different crystal field and flip into the nonmagnetic low spin state. There is thus an As with a different valence (extra charge) near one iron.

One must realize of course, that the magnetically coupled singlet bipolarons are also pairs of two equal electric charges (holes in general for the cuprites, electrons for the Pu and Fe superconductors), which due to their lattice distortion also interact with phonons just as in the BCS model. Thus there will be also a small isotope effect for $O^{18}$, but it has been stated several times, e.g. by inelastic neutron scattering on the iron compounds [27] that phonon interaction in the sense of BCS is not enough to explain the high $T_c$.

In conclusion, it has been shown that the parent materials of high $T_c$ superconductors are antiferromagnets, where long - range magnetic order has been interrupted by 5 – 20% substitution of the magnetic ions by nonmagnetic ions. These nonmagnetic ions have been provoked by chemical doping, but are of the same kind as the magnetic ions, only in another valence state or another spin configuration. The remaining short – range antiferromagnetic clusters or fluctuations will surround such a spin hole with charge as a magnetic polaron. Two such polarons have an attractive interaction and form a boson nonmagnetic bipolaron. This can make a Bose condensation and lead to superconductivity, which has been shown in many papers by Alexandrov and Mott [30, 31]. We could show, that the same mechanism works for all three (Cu, Pu and Fe) high $T_c$ superconducting systems.

## Acknowledgement

Fruitful discussions with B. Batlogg are gratefully acknowledged

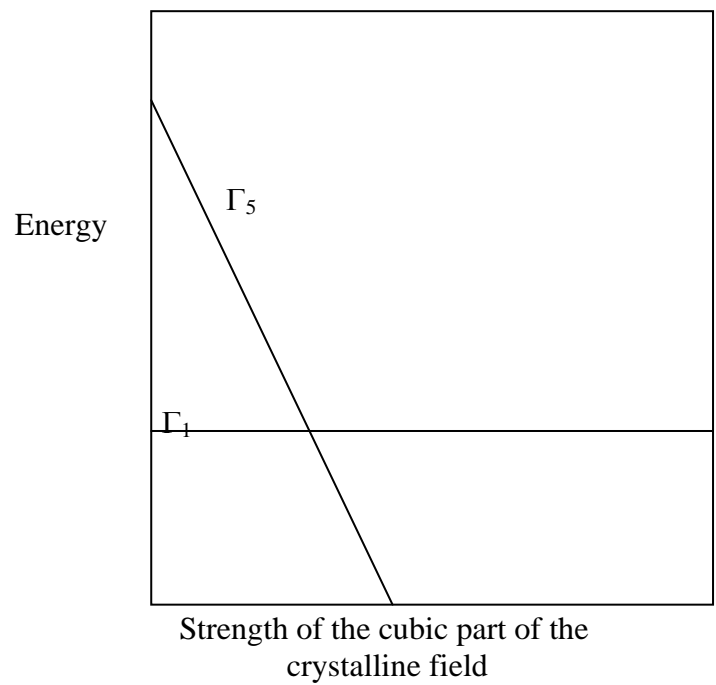

Fig. 1: Schematic representation of a cross-over transition from the high spin ground state $\Gamma_5$ to a low spin ground state $\Gamma_1$ for a 3d$^6$ configuration

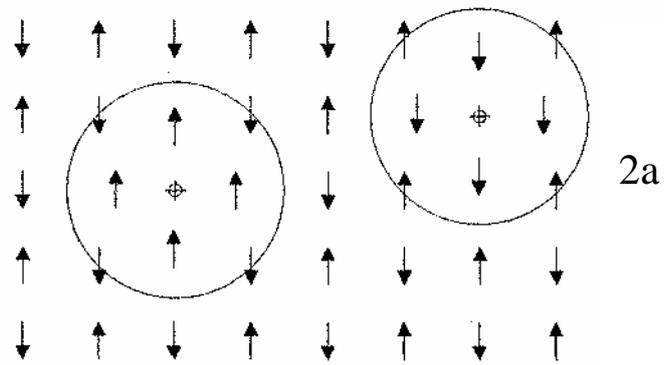

MAGNETIC POLARON

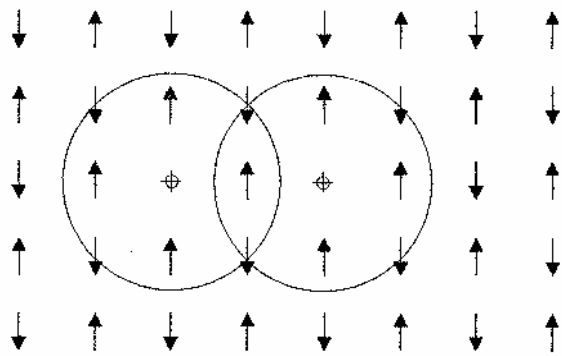

MAGNETIC - BOUND BIPOLARON

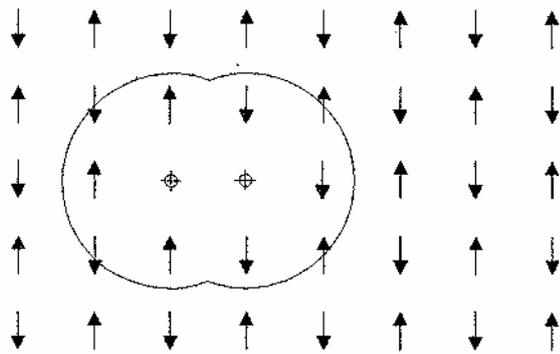

MAGNETIC - BOUND BIPOLARON

Fig. 2. (a) Two separated spin holes $3d^6$ in an antiferromagnetic cluster.
(b) Two spin holes with attractive interaction forming a triplet bipolaron
(c) Two spin holes with attractive interaction forming a nonmagnetic
singlet bipolaron, after [13]